\documentclass[11pt,a4paper]{article}
\usepackage{epsfig,graphicx,amsfonts,amsmath}

\usepackage{cite}
\RequirePackage{mathrsfs}

\newlength{\dinwidth}
\newlength{\dinmargin}
\def\eq#1{{Eq.~(\ref{#1})}}
\newcommand{\Le}{\left(}
\newcommand{\Ra}{\right)}

\newcommand{\beq}{\begin{equation}}
\newcommand{\eeq}{\end{equation}}
\newcommand{\beqar}{\begin{eqnarray}}
\newcommand{\eeqar}{\end{eqnarray}}
\newcommand{\D}{\partial}
\newcommand{\g}{{\rm g}}

\newcommand{\ep}{\varepsilon}

\newcommand{\tv}{\textsl{v}}

\date{}
\begin{document}

\title {{~}\\
{\Large \bf  $S$-matrix and productions amplitudes in high energy QCD }}
\author{ 
{~}\\
{\large 
S.~Bondarenko$^{(1) }$,
S.~Pozdnyakov$^{(1) }$
}\\[7mm]
{\it\normalsize  $^{(1) }$ Physics Department, Ariel University, Ariel 40700, Israel}\\
}

\maketitle
\thispagestyle{empty}

\begin{abstract}
  In this note we discuss a  generalization of the Lipatov's effective action approach, \cite{LipatovEff,LipatovEff1}, for the case of description
of gluon and quark production amplitudes in the quasi-multi-Regge kinematics. Following to \cite{Faddeev}, we define
the S-matrix elements 
of high energy QCD processes in this kinematics and discuss applications of the obtained results. 
\end{abstract}

\section{Introduction}

 $\,\,\,\,\,\,$The Lipatov's effective action approach, \cite{LipatovEff,LipatovEff1}, which introduces in QCD the reggeized gluon fields as new degrees of freedom,
is widely used in the calculation of the high energy QCD scattering amplitudes, \cite{EffAct}. In terms of the reggeon fields, the QCD can be 
formulated as Regge Field Theory (RFT), \cite{Gribov,Our1,Our2,Our3},  
designed to account unitarity corrections to the amplitudes in the forms of reggeon-reggeon interaction vertices, see also \cite{Our4}.
Among the direct use of the approach, it was 
also demonstrated the equivalence of the formalism to the Color Glass Condensate (CGC, see \cite{Venug,Kovner,Hatta1}) approach, see \cite{Our1,Our2,Hetch}. 

 One of the mostly important application of the effective action formalism is the calculation of the production amplitudes in high energy scattering
processes.
Using considered in \cite{EffAct,Cherednikov} Feynman rules, which describes vertices of interactions of particles with reggeons, the different production vertices 
of gluon and quark fields in quasi-multi-Regge kinematics were calculated, see \cite{LipatovEff,Fadin}. 
Nevertheless, discussing the problem of determination of the
S-matrix elements of the theory by receipts proposed in \cite{Faddeev}, we note that we can not use these vertices in the S-matrix elements construction due the fact that 
produced particles in these vertices are off-shell.
Namely, the S-matrix elements in approach of \cite{Faddeev} are defined by the construction of the evolution operator in the way which includes properly defined particle's 
"in" and "out" asymptotic states.
In application to the Lipatov's effective action it means  
that we have to define in the effective action the vertices of transition of reggeon states to asymptotic\footnote{We do not consider problem of confinement here
and define the asymptotic particle's states as solutions of particle's free equations of motion.} gluon's and quarks (antiquark) states (real particles). 
Correspondingly, these new vertices can not be used further for the construction of the unitary corrections in the theory\footnote{There is no need for that in the formulation of the 
approach developed in \cite{LipatovEff,LipatovEff1,Our1,Our2,Our3}.} and they can be understood as the production vertices of real particles in the high energy scattering 
in the quasi-multi-Regge kinematics, for which the Lipatov's action is defined, see \cite{LipatovEff,LipatovEff1,Our1,Our2,Our3,Our4}.
Some part of these vertices also can be interpreted as impact factors of interactions of reggeon fields with 
the gluons or quark (antiquark) fields. In this paper, therefore, we extend the formalism in order to include in the vertices of real particles production, i.e.
the vertices of interaction of the reggeons with asymptotic gluon and quark fields.

  The paper is organized as follows. In the next Section 2 we shortly remind main ideas of the approach and explain the main idea of the formalism's generalization
	for the case of production. Section 3 is about the generalization of the theory for the case of gluon's production amplitudes,
Section 4 is dedicated to the inclusion of quarks in the proposed scheme and last Section 4 is the Conclusion of the paper.

\section{Formalism of Lipatov's effective action}

$\,\,\,\,\,\,$ The effective action, see \cite{LipatovEff,LipatovEff1}, is a non-linear gauge invariant action which correctly reproduces the production of the particles in direct channels at a quasi-multi-Regge kinematics. 
It is written for the local in rapidity interactions of physical gluons in
direct channels inside of some rapidity interval $(y\,-\,\eta/2,y\,+\,\eta/2)$. The interaction between the different clusters of gluons at different though very close rapidities
can be described with the help of reggeized gluon fields
\footnote{We use the Kogut-Soper convention for the light-cone for the light-cone definitions with $x_{\pm}\,=\,\Le x_{0}\,\pm\,x_{3} \Ra/\sqrt{2}$ and $x_{\pm}\,=\,x^{\mp}$\,.} 
$A_{-}$ and $A_{+}$ interacting in crossing channels. Those interaction are non-local in rapidity space. 
This non-local term is not included in the action, the term of interaction between the reggeon fields in the action is local in rapidity and can be considered a 
kind of renormalization term  in the Lagrangian.
The action is gauge invariant and written in the covariant form in terms of gluon field $\textsl{v}$ as
\beq\label{Ef1}
S_{eff}\,=\,-\,\int\,d^{4}\,x\,\Le\,\frac{1}{4}\,F_{\mu \nu}^{a}\,F^{\mu \nu}_{a}\, \,+\,tr\,\left[\,\textsl{v}_{+}\,J^{+}(\textsl{v}_{+})\,-\,A_{+}\,j_{reg}^{+}\,+\,
\,\textsl{v}_{-}\,J^{-}\,(\textsl{v}_{-})\,-\,A_{-}\,j_{reg}^{-}\,\right]\,\Ra\,,
\eeq
where
\beq\label{Ef2}
J^{\pm}(\textsl{v}_{\pm})\,=\,O(x^{\pm}, \textsl{v}_{\pm})\,j_{reg}^{\pm}\,,
\eeq
with $O(x^{\pm}, \textsl{v}_{\pm})$ as some operators, see \cite{LipatovEff,Our1,Our2}, and where
\beq\label{Ef3}
j_{reg\,a}^{\pm}\,=\,\frac{1}{C(R)}\,\D_{i}^{2} A_{a}^{\pm}\,
\eeq
is a reggeon current with $C(R)$ as the eigenvalue of Casimir operator in the representation R with $C(R)\,=\,N$ in the case of adjoint representation.
There are additional kinematical constraints for the reggeon fields
\beq\label{Ef4}
\partial_{-}\,A_{+}\,=\,\partial_{+}\,A_{-}\,=\,0\,,
\eeq
corresponding to the strong-ordering of the Sudakov components in the multi-Regge kinematics,
see \cite{LipatovEff}. Everywhere, as usual,  $\,\partial_{i}\,$ denotes the derivative on transverse coordinates.
Under variation on the gluon fields these currents reproduce the Lipatov's induced currents
\beq\label{Ef5}
\delta\,\Le\,\textsl{v}_{\pm}\, J^{\pm}(\textsl{v}_{\pm})\,\Ra\,=\,\Le \delta\,\textsl{v}_{\pm} \Ra\,j^{ind}_{\mp}(\textsl{v}_{\pm})\,=\,
\Le  \delta\,\textsl{v}_{\pm} \Ra \,j^{\pm}(\textsl{v}_{\pm})\,,
\eeq
with shortness notation $j^{ind}_{\mp}\,=\,j^{\pm}$ introduced, see details in \cite{LipatovEff, Our1}.
This current possesses  a covariant conservation property:
\beq\label{Ef6}
\Le\,D_{\pm}\,j_{\mp}^{ind}(\textsl{v}_{\pm})\,\Ra^{a}\,=\,\Le D_{\pm}\,j^{\pm}(\textsl{v}_{\pm})\Ra^{a}\,=\,0\,.
\eeq
It was shown in \cite{Our1}, that fixing the LO value of classical gluon field in solutions as
\beq\label{Ef8}
\textsl{v}_{\pm\,cl}\,=\,A_{\pm}
\eeq
and applying the self-consistency conditions on the classical solutions, see \cite{Our1}, the currents in \eq{Ef1} Lagrangian can be reproduced directly in form of \eq{Ef2} without  any additional conditions.
Applying  the light-cone gauge $\textsl{v}_{-}\,=\,0$ and solving  the equations of motion,  the general expressions for the gluon fields can be written as
\beq\label{Ef9}
v_{i}^{a}\,\rightarrow\,v^{a}_{i\,cl}\,+\,\ep_{i}^{a}\,,\,\,\,\,v_{+}^{a}\,\rightarrow\,v_{+\,cl}^{a}\,+\,\ep_{+}^{a}\,,
\eeq
where the integration on fluctuations around the classical solutions provides loop corrections to the "net" contribution which is based on the classical solutions only.

 The use of the formalism in the form defined above, nevertheless, provides only quasi-elastic contribution to the scattering amplitudes in the 
quasi-multi-Regge high energy kinematics. In order to generalize the approach for the description of the production of the real particles in the scattering
processes we need to include  the real particle asymptotic states in \eq{Ef9} in the consistent manner, this task is considered in the following Sections.

\section{S-matrix for the gluon production amplitudes}

 In order to compute the production amplitudes in the quasi-multi-Regge kinematics, the classical solutions obtained and used in \cite{Our1,Our2,Our3} must be modified:
\beq\label{GC1}
\textsl{v}^{\,\mu}_{cl}\,\rightarrow\,\tv^{\,\mu}_{cl}\,+\,\tv^{\,\mu}_{f}
\eeq
with $\tv^{\mu}_{f}$ as solutions of equations of motion for free gluon field. 
The full gluon fields, therefore, will acquire the following form:
\beq\label{GC11}
\textsl{v}^{\,\mu}\,=\tv^{\,\mu}_{cl}\,+\,\tv^{\,\mu}_{f}\,+\,\xi\,
\eeq 
with $\xi$ as fluctuations above the found classical solutions. The $\tv^{\,\mu}_{cl}\,$ fields were calculated in \cite{Our3} to NNLO precision,
here we consider the equations of motion for $\tv^{\,\mu}_{f}$ gluon fields 
\beq\label{GC2}
\Le\,D_{\mu}\,G^{\mu \nu}\,\Ra_{a}\,=\,\partial_{\mu}\,G_{a}^{\mu \nu}\,+\,g\,f_{abc}\textsl{v}_{\mu}^{b}\,G^{c\,\mu \nu}\,=\,0\,,
\eeq
to LO precision. These equations provide the well known expressions for $\tv_{+}$ and $\tv_{\bot}$ components of the gluon field, i.e. 
the first equation provides a constraint of the theory in light-cone gauge, it gives the  expression of the $\tv_{+}$
component of the field in terms of the transverse components:
\beq\label{GC3}
\tv_{a\,+\,f}\,=\,-\,\partial^{-1}_{-}\,\partial_{i}\,\tv^{i}_{a\,f}\,,
\eeq
see \cite{LightCone}.
The second equation of motion gives the equation of motion for the dynamical $\tv_{i}$ fields:
\beq\label{GC4}
\Le\,2\,\partial_{+}\,\partial_{-}\,-\,\partial_{\bot}^{2}\,\Ra\,\tv_{i\,a\,f}\,=\,\Box\,\tv_{i\,a\,f}\,=\,0\,.
\eeq
The third equation can be considered as check of the consistency of the found solutions, it is equal to zero by construction, see \cite{Our3}. 
We see also, that the Lorentz condition for the free gluon field
\beq\label{GC5}
\partial^{\,\mu}\,\tv_{\mu\,f}\,=\,0
\eeq
is satisfied in this case as an operator equation\footnote{ The momentum representation of the dynamical fields in terms of annihilation-creation operators is also well-known:
\beqar\label{GC6}
\tv^{a}_{i\,f}(x^{+},\,x^{-},\,x_{\bot})\,& = &\,\frac{1}{(2\pi)^{3/2}}\,\int_{k_{-}\, > \,0}\,\frac{dk_{-}\,d^{2} k_{i}}{\sqrt{2\,k_{-}}}\,\Le\,
\textsl{a}_{i}^{a}(x^{+},\,k_{-},\,k_{\bot})\,e^{-\imath\,k_{-}\,x^{-}\,+\,\imath\,k_{\bot}\,x_{\bot}}\,+\,\right.\nonumber \\
&+&\,\left.\textsl{a}_{i}^{\dag\,a}(x^{+},\,k_{-},\,k_{\bot})\,e^{\imath\,k_{-}\,x^{-}\,-\,\imath\,k_{\bot}\,x_{\bot}}\,\Ra\,\nonumber,
\eeqar
where at asymptotic times $x^{+}\,\rightarrow\,\pm\infty$ we have:
\beq\label{GC7}\nonumber
\textsl{a}_{i}^{a}(x^{+},\,k_{-},\,k_{\bot})\,\rightarrow\,\textsl{a}_{i}^{a}(\,k_{-},\,k_{\bot})\,e^{-\imath\,k_{\bot}^{2}\,x^{+}\,/\,2\,k_{-}}\,+\,
\tilde{\textsl{a}}_{i}^{a}(x^{+},\,k_{-},\,k_{\bot})\,,\,\,\,\,\tilde{\textsl{a}}_{i}^{a}(x^{+},\,k_{-},\,k_{\bot})\,\rightarrow\,0\,,\,\,\,\,x^{+}\,\rightarrow\,-\,\infty
\eeq
and
\beq\label{GC8}\nonumber
\textsl{a}_{i}^{\dag\,a}(x^{+},\,k_{-},\,k_{\bot})\,\rightarrow\,\textsl{a}_{i}^{\dag\,a}(\,k_{-},\,k_{\bot})\,e^{\imath\,k_{\bot}^{2}\,x^{+}\,/\,2\,k_{-}}\,+\,
\tilde{\textsl{a}}_{i}^{\dag\,a}(x^{+},\,k_{-},\,k_{\bot})\,,\,\,\,\,\tilde{\textsl{a}}_{i}^{\dag\,a}(x^{+},\,k_{-},\,k_{\bot})\,\rightarrow\,0\,,\,\,\,\,x^{+}\,\rightarrow\,\infty\,,
\eeq
see \cite{Faddeev,LightCone}.}.

  With the new classical solution obtained, following to the approach of \cite{LipatovEff,Our1,Our2}, the effective action becomes to be a functional of the 
Reggeon and free gluon fields: $\Gamma(A_{+},\,A_{-},\,\tv_{f})$. 
The $S$-matrix generating functional of this sector of the approach consists with free gluon fields, it can be written as
\beq\label{GC19}
S[\tv_{\bot f}]\,=\,\int\,D A\,e^{\imath\,\Gamma(A_{\pm},\,\tv_{\bot f})}\,=\,\int\,D A\,\int\,D \xi\,e^{\imath\,S(\tv)}\,,
\eeq
see \cite{Faddeev}.
Taking functional derivatives with respect to these fields from the $\Gamma(A_{\pm},\,\tv_{\bot f})$ effective action, any production vertex of interest can be calculated by the 
straightforward way. Namely, vertex of interaction of $n$ $A_{+}$ fields with $m$ $A_{-}$ fields and $k$ $\tv_{\bot}$ fields we can write as 
\beq\label{GC81}
\Le\, K_{a_1\cdots a_{n}}^{b_1\cdots b_{m}}\,\Ra^{c_{1}\cdots c_{k}}\,=\,\Le\,
\frac{\delta^{n\,+\,m\,+\,k\,}\,\Gamma(A_{+},\,A_{-},\,\tv_{f})}
{\delta A_{+}^{a_1}\cdots \delta A_{+}^{a_1}\,\delta A_{-}^{b_1}\cdots \delta A_{-}^{b_m}\,\delta \tv_{\bot\,f}^{\,c_1}\cdots \delta\tv_{\bot\,f}^{\,c_k}}\,\Ra_{A_{\pm}\,=\,\tv_{\bot\,f}\,=\,0}.
\eeq
Additionally, considering the impact factor of the theory as effective vertex of the 
interaction of the reggeons of one kind with the real particles, we can correspondingly define the 
impact factors of interaction of $n$ reggeon fields with $m$ gluons as
\beq\label{GC82}
 \tilde{K}_{a_1\cdots a_{n}}^{b_1\cdots b_{m}}\,=\,\Le\,
\frac{\delta^{n\,+\,m\,}\,\Gamma(A_{+},\,A_{-},\,\tv_{f})}
{\delta A_{\pm}^{a_1}\cdots \delta A_{\pm}^{a_n}\,\delta \tv_{\bot\,f}^{\,b_1}\cdots \delta\tv_{\bot\,f}^{\,b_m}}\,\Ra_{A_{\pm}\,=\,\tv_{\bot\,f}\,=\,0}\,,\,\,\,m\,=\,2,\,3,\,\cdots\,,
\eeq
In this paper we do not calculate any vertices of these kind, restricting only by the formalism's clarification. The task of the calculation of the expression for the
different vertices will be considered in a separate publication.

\section{S-matrix for the quark production amplitudes}

 Considering the fermionic part of the QCD Lagrangian, we follow to the usual light-cone decomposition of the quark field , see \cite{LightCone}.
Introducing projector operators
\beq\label{QS1}
\Lambda^{\pm}\,=\,\frac{\gamma^{0}\,\gamma^{\pm}}{\sqrt{2}}\,=\,\frac{\gamma^{\mp}\,\gamma^{\pm}}{2}\,,\,\,\,\gamma^{\pm}\,=\,\frac{\gamma^{0}\,\pm\,\gamma^{3}}{\sqrt{2}}\,,
\eeq
we project out the fermion field obtaining two two-component spinors:
\beq\label{QS2}
\psi_{\pm}\,=\,\Lambda^{\pm}\,\psi\,.
\eeq
The Lagrangian of the interacting of fermion fields in the case of the decomposition acquires the following form:
\beqar
L_{Q}\,&= &\,\bar{\psi}_{-}\,\Le\, \imath\,\gamma^{\bot}\,D_{\bot}\,-\,m\,\Ra\,\psi_{+}\,+\,\bar{\psi}_{+}\,\Le\, \imath\,\gamma^{\bot}\,D_{\bot}\,-\,m\,\Ra\,\psi_{-}\,+\,
\imath\,\sqrt{2}\,\bar{\psi}_{-}\,\gamma^{0}\,\partial_{-}\,\psi_{-}\,+
\nonumber \\
&+&\,\imath\,\sqrt{2}\,\bar{\psi}_{+}\,\gamma^{0}\,D_{+}\,\psi_{+}\,\label{QS3}
\eeqar 
with covariant derivative defined for the gluon field in the fundamental representation:
\beq\label{QS4}
D_{\mu}\,=\,\partial_{\mu}\,-\,g\,\tv_{\mu}\,=\,\partial_{\mu}\,-\,\imath\,g\,\tv_{\mu}^{\,a}\,T^{a}\,.
\eeq
The equations of motion for the fermion fields contains a constraint for the $\psi_{-}$ field\footnote{
In the Lagrangian path integral formalism the constraint can be understood as possibility to fully integrate out the $\psi_{-}$ field from the path integral without
quantum corrections in the final result, i.e. as
$$
N^{-1}\,\int\,D\psi_{-}\,D\psi_{-}^{\dag}\,e^{\imath S_{Q}[\psi_{+},\,\psi_{-}]}\,=\,N^{'\,-1}\,e^{\imath S_{Q}[\psi_{+}]}
$$
where in the l.h.s of the expression the action is defined with the help of \eq{QS3} Lagrangian whereas in the r.h.s of the equality the action is defined already
with the Lagrangian from \eq{QS6}. Correspondingly, the presence of the Lipatov's effective currents in \eq{Ef1} Lagrangian of the gluon sector of the approach, 
does not allow to integrate out the $\tv_{+}$ gluon field as it usually done in the ordinary light-cone formulation of the QCD. We also note, that 
in the theory with reggeized quarks included, see \cite{Nefedov}, the constraint \eq{QS5} will not be hold anymore as well. We grateful to M.Zubkov who pointed out 
this formulation of the constraint in the formalism. }:
\beq\label{QS5}
\psi_{-\,x}\,=\,\frac{\imath}{\sqrt{2}}\,\gamma^{0}\,\partial_{-\, x y}^{-1}\,\Le\,\imath\,\gamma^{\bot}\,D_{\bot}\,-\,m\,\Ra_{y}\,\psi_{+\,y}
\eeq
which leads to the following expression of the fermion Lagrangian after the inserting of the constraint back into the \eq{QS3}:
\beqar
L_{Q}\, &=& \,\imath\,\sqrt{2}\,\bar{\psi}_{+}\,\gamma^{0}\,D_{+}\,\psi_{+}\,+\,\frac{\imath}{\sqrt{2}}\,\bar{\psi}_{+}\,
\Le\, \imath\,\gamma^{\bot}\,D_{\bot}\,-\,m\,\Ra\,\gamma^{0}\,\partial_{-}^{-1}\,\Le\, \imath\,\gamma^{\bot}\,D_{\bot}\,-\,m\,\Ra\,\psi_{+}\,=\,\nonumber\\
&=&\,\imath\,\sqrt{2}\,\psi_{+}^{\dag}\,D_{+}\,\psi_{+}\,-\,\frac{\imath}{\sqrt{2}}\,\psi_{+}^{\dag}\,
\Le\, \imath\,\gamma^{\bot}\,D_{\bot}\,+\,m\,\Ra\,\partial_{-}^{-1}\,\Le\, \imath\,\gamma^{\bot}\,D_{\bot}\,-\,m\,\Ra\,\psi_{+}\,\label{QS6}.
\eeqar
The free, asymptotic quark fields we can define as solutions of equations of motion obtained from the free quarks Lagrangian
\beq\label{QS7}
L_{Q}^{f}\,=\,\imath\,\sqrt{2}\,\psi_{+}^{\dag}\,\partial_{+}\,\psi_{+}\,-\,\frac{\imath}{\sqrt{2}}\,\psi_{+}^{\dag}\,
\Le\, \imath\,\gamma^{\bot}\,\partial_{\bot}\,+\,m\,\Ra\,\partial_{-}^{-1}\,\Le\, \imath\,\gamma^{\bot}\,\partial_{\bot}\,-\,m\,\Ra\,\psi_{+}\,
\eeq
and which reads as
\beq\label{QS8}
\Le\,2\,\partial_{+}\,-\,\Le\, \imath\,\gamma^{\bot}\,\partial_{\bot}\,+\,m\,\Ra\,\partial_{-}^{-1}\,\Le\, \imath\,\gamma^{\bot}\,\partial_{\bot}\,-\,m\,\Ra\,\Ra\,\psi_{+\,f}\,=\,0\,,
\eeq
or as\footnote{In the momentum space, the asymptotic behavior of the quark field is similar to the asymptotic behavior of the gluon field, see footnote in the previous Section.  }
\beq\label{QS81}
\Le\,2\,\partial_{+}\,\partial_{-}-\,\D_{\bot}^{2}\,+\,m^{2}\,\Ra\,\psi_{+\,f}\,=\,0\,.
\eeq
Correspondingly, the propagator of the free quark fields we define trough the following equation
\beq\label{QS9}
\frac{\imath}{\sqrt{2}}\,\Le\,2\,\partial_{+}\,-\,\Le\, \imath\,\gamma^{\bot}\,\partial_{\bot}\,+\,m\,\Ra\,\partial_{-}^{-1}\,\Le\, \imath\,\gamma^{\bot}\,\partial_{\bot}\,-\,m\,\Ra\,\Ra_{x}\,
G_{0}(x,y)\,=\,\delta^{4}(x\,-\,y)\,,
\eeq
with all group's indexes suppressed in the definition. The solution of \eq{QS9} is simple:
\beq\label{QS10}
G_{0}^{\,i j}(x,y)\,=\,\sqrt{2}\,\delta^{\,i j}\,\int\,\frac{d^{4} p\,}{(2\,\pi)^{4}}\frac{p_{-}\,e^{-\imath\,p\,(x-y)}}{\Le\,p^{2}\,-\,m^{2}\,+\,\imath\,\varepsilon\,  \Ra}\,.
\eeq
Now, following to our approach, we expand the quark fields as sum of free solution and additional quark's field fluctuation above it:
\beq\label{QS11}
\psi_{+}\,=\,\psi_{+\,f}\,+\,\epsilon_{+}\,.
\eeq
Inserting this expression back in \eq{QS6}, we obtain for the quark's Lagrangian:
\beqar\label{QS13}
L_{Q}\,&=&\,\frac{\imath}{\sqrt{2}}\,\epsilon_{+}^{\dag}\,\Le\,2\,D_{+}\,-\,\Le\, \imath\,\gamma^{\bot}\,D_{\bot}\,+\,
m\,\Ra\,\partial_{-}^{-1}\,\Le\, \imath\,\gamma^{\bot}\,D_{\bot}\,-\,m\,\Ra\,\Ra\,\epsilon_{+}\,+\,\nonumber \\
&+&\,\epsilon_{+}^{\dag}\,J_{0}(\tv,\,\psi_{+\,f})\,+\,J^{\dag}(\tv,\,\psi_{+\,f}^{\dag})\,\epsilon_{+}\,+\,\psi_{+\,f}^{\dag}\,M_{0}(\tv)\,\psi_{+\,f}\,,
\eeqar
with the currents
\beqar\label{QS14}
J_{0}(\tv,\,\psi_{+\,f})\,& = &\frac{1}{\sqrt{2}}\,\Le\,2\,g\,\tv_{+}^{a}\,t^{a}\,
-\,\imath\,\Le\, g\,\gamma^{\bot}\,\tv_{\bot}^{a}\,T^{a} \,\Ra\,\partial_{-}^{-1}\,\Le\,\imath\,\gamma^{\bot}\,\partial_{\bot}\,-\,m\,\Ra\,-\,\right.\nonumber \\
&-&\left.\imath\Le\imath\gamma^{\bot}\partial_{\bot}+m\,\Ra\,\partial_{-}^{-1}\,
\Le\g\gamma^{\bot}\tv_{\bot}^{a}\,T^{a}\Ra-
 \imath\Le g \gamma^{\bot} \tv_{\bot}^{a}\,T^{a} \Ra\partial_{-}^{-1}
\Le g \gamma^{\bot} \tv_{\bot}^{b}\,T^{b} \Ra\Ra\psi_{+\,f}\,
\eeqar
\beqar\label{QS141}
J_{0}^{\dag}(\tv,\,\psi_{+\,f})\,& = &\frac{1}{\sqrt{2}}\,\psi_{+\,f}^{\dag}\,\Le\,2\,g\,\tv_{+}^{a}\,t^{a}\,
-\,\imath\,\Le\, g\,\gamma^{\bot}\,\tv_{\bot}^{a}\,T^{a} \,\Ra\,\partial_{-}^{-1}\,\Le\,\imath\,\gamma^{\bot}\,\partial_{\bot}\,-\,m\,\Ra\,-\,\right.\nonumber \\
&-&\left.\imath\Le\imath\gamma^{\bot}\partial_{\bot}+m\,\Ra\,\partial_{-}^{-1}\,
\Le g\gamma^{\bot}\tv_{\bot}^{a}\,T^{a}\Ra-
 \imath\Le g \gamma^{\bot} \tv_{\bot}^{a}\,T^{a} \Ra\partial_{-}^{-1}
\Le g \gamma^{\bot} \tv_{\bot}^{b}\,T^{b} \Ra\Ra\,,
\eeqar
and with the bare kernel of interaction of the free quark fields with the gluon fields:
\beqar\label{QS15}
M_{0}(\tv)\,&=&\,\frac{1}{\sqrt{2}}\,\Le\,2
g\,\tv_{+}^{a}\,t^{a}\,-\,
\imath\,\Le\, g\,\gamma^{\bot}\,\tv_{\bot}^{a}\,T^{a} \,\Ra\,\partial_{-}^{-1}\,\Le\,\imath\,\gamma^{\bot}\,\partial_{\bot}\,-\,m\,\Ra\,-\,\right.\nonumber \\
&-&\,\left.\imath\,\Le\,\imath\,\gamma^{\bot}\,\partial_{\bot}\,+\,m\,\Ra\,\partial_{-}^{-1}\,
\Le\, g\,\gamma^{\bot}\,\tv_{\bot}^{a}\,T^{a} \,\Ra\,-\,
\imath\,\Le\, g\,\gamma^{\bot}\,\tv_{\bot}^{a}\,T^{a} \,\Ra\,\partial_{-}^{-1}\,
\Le\, g\,\gamma^{\bot}\,\tv_{\bot}^{b}\,T^{b} \,\Ra\,\Ra\,.
\eeqar
Using equations of motion for the free $\psi_{+}$ fields 
\beq\label{QS16}
\partial_{-}\,\Le\,\imath\,\gamma^{\bot}\,\partial_{\bot}\,-\,m\,\Ra\,\psi_{+\,f}\,=\,2\,\Le\,\imath\,\gamma^{\bot}\,\partial_{\bot}\,+\,m\,\Ra^{-1}\,\partial_{+}\,\psi_{+\,f}\,
\eeq
and
\beq\label{QS17}
\partial_{-}\,\Le\,\imath\,\gamma^{\bot}\,\partial_{\bot}\,+\,m\,\Ra\,\,\psi_{+\,f}^{\dag}\,=\,-\,2\,\Le\,\imath\,\gamma^{\bot}\,\partial_{\bot}\,-\,m\,\Ra^{-1}\,
\partial_{+}\,\psi_{+\,f}^{\dag}\,,
\eeq
the \eq{QS15} expression can be rewritten also as
\beqar\label{QS18}
M_{0}(\tv)\,&=&\,\frac{1}{\sqrt{2}}\,\Le\,2
g\,\tv_{+}^{a}\,t^{a}\,-\,2\,
\imath\,\Le\, g\,\gamma^{\bot}\,\tv_{\bot}^{a}\,T^{a} \,\Ra\,\Le\,\imath\,\gamma^{\bot}\,\partial_{\bot}\,+\,m\,\Ra^{-1}\,\partial_{+}\,-\,\right.\nonumber \\
&-&\left.
2\imath\,\partial_{+}\,\Le\imath\,\gamma^{\bot}\,\partial_{\bot}-m\,\Ra^{-1}\,\partial_{-}^{-1}
\Le g\,\gamma^{\bot}\,\tv_{\bot}^{a}\,T^{a} \Ra-
\imath\,\Le g\,\gamma^{\bot}\,\tv_{\bot}^{a}\,T^{a} \Ra\,\partial_{-}^{-1}
\Le g\,\gamma^{\bot}\,\tv_{\bot}^{b}\,T^{b} \Ra\Ra\,.
\eeqar
Now, the QCD  $S$-matrix generating functional of the theory consists with free gluon and quark fields we can write as
\beq\label{QS19}
S[\tv_{\bot f},\,\psi_{+\,f},\,\psi^{\dag}_{+\,f}]\,=\,\int\,D A\,e^{\imath\,\Gamma(A_{\pm},\,\tv_{\bot f},\,\psi_{+\,f},\,\psi^{\dag}_{+\,f})}\,=\,
\int\,D A\,\int\,D \xi\,D \epsilon_{+}\,e^{\imath\,S(\tv ,\,\psi)}\,,
\eeq
with $\Gamma(A_{\pm},\,\tv_{\bot f},\,\psi_{+},\,\psi^{\dag}_{+})$ as an effective action of the theory.
Correspondingly,
taking functional derivatives of $\Gamma$ with respect to these fields, we can calculate any vertex vertex of the theory which consists with the fields, i.e. similarly to \eq{GC8} definition, 
vertex of interaction of $n$ $A_{+}$ fields, $m$ $A_{-}$ fields,  $k$ $\tv_{\bot}$ fields, $p$ $\psi_{+}$ and $h$ $\psi_{+}^{\dag}$ fields  we can write as:
\beq\label{QS20}
\Le\, K_{a_1\cdots a_{n}}^{b_1\cdots b_{m}}\,\Ra^{c_{1}\cdots c_{k}}_{p\,;\,h}\,=\,\Le\,
\frac{\delta^{n\,+\,m\,+\,k\,+\,p\,+\,h}\,\Gamma(A_{\pm},\,\tv_{\bot f},\,\psi_{+\,f},\,\psi^{\dag}_{+\,f})}
{\delta A_{+}^{a_1}\cdots \delta A_{+}^{a_n}\,\delta A_{-}^{b_1}\cdots \delta A_{-}^{b_m}\,
\delta \tv_{\bot\,f}^{\,c_1}\cdots \delta\tv_{\bot\,f}^{\,c_k}\,
\delta \psi_{+ \,f}^{1}\cdots \delta\psi_{+ \,f}^{p}\,
\delta \psi_{+ \,f}^{1\,\dag}\cdots \delta\psi_{+\,f}^{h\,\dag}}\,\Ra
\eeq
where as usual we take all the fields equal to zero after the derivatives taken.
Correspondingly, the impact factor of interaction of $n$ reggeon fields of the same kind with $k$ free gluon field and $2\,p$ quark and anti-quark fields we define as
\beq\label{QS201}
\Le\, \tilde{K}_{a_1\cdots a_{n}}\,\Ra^{c_{1}\cdots c_{k}}_{p\,;\,p}\,=\,\Le\,
\frac{\delta^{n\,+\,k\,+\,2\,p\,}\,\Gamma(A_{\pm},\,\tv_{\bot f},\,\psi_{+\,f},\,\psi^{\dag}_{+\,f})}
{\delta A_{\pm}^{a_1}\cdots \delta A_{\pm}^{a_n}\,
\delta \tv_{\bot\,f}^{\,c_1}\cdots \delta\tv_{\bot\,f}^{\,c_k}\,
\delta \psi_{+ \,f}^{1}\cdots \delta\psi_{+ \,f}^{p}\,
\delta \psi_{+ \,f}^{1\,\dag}\cdots \delta\psi_{+ \,f}^{p\,\dag}}\,\Ra_{A=\tv=\psi=\psi^{\dag}=0}\,,
\eeq
the calculation of these vertices we also postpone for a separate publication.

\section{Conclusion}

 In this note we extended the formalism of the Lipatov's effective action for the case of calculation of production amplitudes in 
quasi-multi-Regge kinematics at high energy. The main ideas behind the generalization of the theory were proposed already in \cite{LipatovEff}, 
in the paper we explored these proposals.

 In the previous calculations, \cite{Our1,Our2,Our3}, the considered amplitudes are described the quasi-elastic processes of the target-projectile
interactions at high energies. In the Lipatov's formalism these amplitudes are represented by reggeons t-channel exchanges, where each reggeon field itself includes 
cluster of interacting locally in rapidity real particles. Nevertheless, asking what is the production vertices of real particles in the formalism
we reveal the need to generalize the approach in the way when the vertices of the interactions of the  reggeon fields with real particles will be included in the 
calculations in the consistent way.  Therefore, the main idea of this extension of the formalism is given by \eq{GC11} and \eq{QS11}, we redefine the gluon and quark fields in the way
that all other corresponding calculations can be performed similarly to the calculations done in \cite{Our1}-\cite{Our3}, whereas the obtaining effective theory will include now
the vertices of the production of any number of real particles in the reggeon-reggeon interactions.
In general, 
calculating the Lipatov's effective action once with the precision determined by the QFT methods, we obtain as the result 
the possibility to calculate any kind of vertices in the framework of the RFT. It seems to be an important advantage of the Lipatov's effective action formalism.

 Introducing in the effective action the real fields as correct "in" and "out" QCD asymptotic fields , see \eq{GC4} and \eq{QS8}, we note, that in accordance with the
approach of \cite{Faddeev}, we correspondingly determine correctly the S-matrix of the high energy scattering processes. Indeed, after the integration on the reggeon fields, see \eq{GC19} and \eq{QS19},
we can expand the the S-matrix functional in \eq{GC19} and \eq{QS19} in respect to the free asymptotics felds $\phi_{f}\,=\,\tv_{\bot\,f},\,\psi_{+\,f},\,\psi_{+\,f}^{\dag}$ as:
\beq\label{Con1}
S[\phi_{f}]\,=\,\sum_{n}\,\frac{1}{n!}\,\int\,S_{n}(x_1,\,\cdots,\,x_{n})\,\phi_{f}(x_1)\,\cdots\,\phi_{f}(x_{n})\,dx_{1}\,\cdots\,dx_{n}\,
\eeq
obtaining S-matrix coefficient function $S_{n}$ without the use of reduction formulae. This is another interesting property of the formulated RFT, which
requires a consideration of the RFT as QFT with an integration over the reggeon field performed. So far, the integration of the reggeon loops, required
as part of the unitary corrections in the RFT, is a difficult task, see \cite{RegLoops}.

In conclusion we note, that the developed formalism is a powerful tool for the calculation of any kind amplitudes and unitary corrections in the
QCD high-energy scattering. Our next step, therefore, it is a calculation of the vertices of interests in the approach 
and use them for the further theory's development and data description.

\section{Acknowledgments}

 The idea of this paper was inspired by numerous discussions with L.Lipatov. 
Both authors are grateful to M.Zubkov and A.Prygarin for useful discussions.


\newpage
\begin{thebibliography}{99}


\bibitem{LipatovEff}
L.~N.~Lipatov,
  Nucl. Phys. B {\bf 452}, 369 (1995); Phys. Rept.  {\bf 286}, (1997) 131.
	
\bibitem{LipatovEff1}
L.~N.~Lipatov,
  Subnucl. Ser.  {\bf 49},(2013) 131;
  Int. J. Mod. Phys. Conf. Ser.  {\bf 39},  (2015) 1560082;
	Int. J. Mod. Phys. A {\bf 31}, no. 28/29,  (2016) 1645011;
	EPJ Web Conf.  {\bf 125}, (2016) 01010.
	
\bibitem{Faddeev}	
L.D. Faddeev, A.A. Slavnov,
"Gauge Fields. Introduction to Quantum Theory",
The Benjamin Cummings Publishing Company, 1980.


\bibitem{EffAct}
L.~N.~Lipatov,
  Nucl.\ Phys.\ Proc.\ Suppl.\  {\bf 99A}, (2001) 175;
	 M.~A.~Braun and M.~I.~Vyazovsky,
  Eur.\ Phys.\ J.\ C {\bf 51}, (2007) 103;
	 M.~A.~Braun, M.~Y.~Salykin and M.~I.~Vyazovsky,
  Eur.\ Phys.\ J.\ C {\bf 65}, (2010) 385;
	M.~A.~Braun, L.~N.~Lipatov, M.~Y.~Salykin and M.~I.~Vyazovsky,
  Eur.\ Phys.\ J.\ C {\bf 71}, (2011) 1639;
	 M.~A.~Braun, M.~Y.~Salykin and M.~I.~Vyazovsky,
  Eur.\ Phys.\ J.\ C {\bf 72}, 1864 (2012);
	M.~Hentschinski and A.~Sabio Vera,
  Phys.\ Rev.\ D {\bf 85}, 056006 (2012);
	M.~A.~Braun, M.~Y.~Salykin, S.~S.~Pozdnyakov and M.~I.~Vyazovsky,
  Eur.\ Phys.\ J.\ C {\bf 72}, (2012) 2223;
	J.~Bartels, L.~N.~Lipatov and G.~P.~Vacca,
  Phys.\ Rev.\ D {\bf 86},  (2012) 105045;
	M.~A.~Braun, S.~S.~Pozdnyakov, M.~Y.~Salykin and M.~I.~Vyazovsky,
  Eur.\ Phys.\ J.\ C {\bf 73}, no. 9, (2013) 2572;
	G.~Chachamis, M.~Hentschinski, J.~D.~Madrigal Martínez and A.~Sabio Vera,
  Phys.\ Part.\ Nucl.\  {\bf 45}, no. 4, (2014) 788;
	 M.~A.~Braun,
  Eur.\ Phys.\ J.\ C {\bf 75} (2015) no.7,  298;
	M.~A.~Braun and M.~I.~Vyazovsky,
  Phys.\ Rev.\ D {\bf 93} (2016) no.6,  065026;
	 M.~A.~Braun,
  Eur.\ Phys.\ J.\ C {\bf 77} (2017) no.5,  279;
	M.~A.~Braun and M.~Y.~Salykin,
  Eur.\ Phys.\ J.\ C {\bf 77} (2017) no.7,  498.

\bibitem{Gribov}
V. N. Gribov, Sov. Phys. JETP 26 (1968) 414.
	
\bibitem{Our1} 
  S.~Bondarenko, L.~Lipatov and A.~Prygarin,
 Eur.\ Phys.\ J.\ C {\bf 77} (2017) no.8,  527.
	

\bibitem{Our2} 
S.~Bondarenko, L.~Lipatov, S.~Pozdnyakov and A.~Prygarin,
  Eur.\ Phys.\ J.\ C {\bf 77} (2017) no. 9, 630. 
	
\bibitem{Our3}
  S.~Bondarenko and S.~S.~Pozdnyakov,
  arXiv:1802.05508 [hep-ph].	
	
\bibitem{Our4}	
S.~Bondarenko and M.~A.~Zubkov,
  arXiv:1801.08066 [hep-ph].	
	
\bibitem{Venug}
L.~McLerran and R.~Venugopalan, {\it{Phys.\ Rev.}}\ {\bf{D49}} (1994), 2233;
\ {\bf{D49}} (1994), 3352.


\bibitem{Kovner}
J.~Jalilian-Marian, A.~Kovner, L.~McLerran and H.~Weigert, 
{\it Phys.Rev.} {\bf D55},
(1997) 5414;
J.~Jalilian-Marian, A.~Kovner, A.~Leonidov and H.~Weigert,
  Nucl.\ Phys.\ B {\bf 504}, (1997) 415;
	J.~Jalilian-Marian, A.~Kovner, A.~Leonidov and H.~Weigert,
  Phys.\ Rev.\ D {\bf 59},(1998) 014014 ;
	E.~Iancu, A.~Leonidov and L.~D.~McLerran,
  Nucl.\ Phys.\ A {\bf 692}, (2001) 583;
	E.~Iancu, A.~Leonidov and L.~D.~McLerran,
  Phys.\ Lett.\ B {\bf 510}, (2001) 133;
	E.~Ferreiro, E.~Iancu, A.~Leonidov and L.~McLerran,
  Nucl.\ Phys.\ A {\bf 703}, (2002) 489;
K.~Roy and R.~Venugopalan,
  arXiv:1802.09550 [hep-ph].	

\bibitem{Hatta1}
I.~Balitsky,
  Phys.\ Rev.\ D {\bf 72},  (2005) 074027;
Y.~Hatta,
  Nucl.\ Phys.\ A {\bf 768}, (2006) 222;
	Y.~Hatta, E.~Iancu, L.~McLerran, A.~Stasto and D.~N.~Triantafyllopoulos,
  Nucl.\ Phys.\ A {\bf 764},(2006) 423.	
Y.~Hatta,
  Nucl.\ Phys.\ A {\bf 781}, (2007) 104.	
	
\bibitem{Hetch}	
M.~Hentschinski,
  arXiv:1802.06755 [hep-ph];

\bibitem{Fadin}	
V.~S.~Fadin and R.~Fiore,
  Phys.\ Lett.\ B {\bf 294} (1992) 286;
V.~S.~Fadin, R.~Fiore and A.~Quartarolo,
  Phys.\ Rev.\ D {\bf 50} (1994) 2265;
		 V.~S.~Fadin, R.~Fiore and A.~Quartarolo,
  Phys.\ Rev.\ D {\bf 50} (1994) 5893;
V.~S.~Fadin, R.~Fiore and M.~I.~Kotsky,
  Phys.\ Lett.\ B {\bf 389} (1996) 737;	
V.~S.~Fadin, R.~Fiore and A.~Papa,
  Phys.\ Rev.\ D {\bf 63} (2001) 034001;	
V.~S.~Fadin and R.~Fiore,
  Phys.\ Rev.\ D {\bf 64} (2001) 114012
	V.~S.~Fadin, M.~G.~Kozlov and A.~V.~Reznichenko,
  Phys.\ Atom.\ Nucl.\  {\bf 67} (2004) 359
   [Yad.\ Fiz.\  {\bf 67} (2004) 377];
M.~G.~Kozlov, A.~V.~Reznichenko and V.~S.~Fadin,
  Phys.\ Atom.\ Nucl.\  {\bf 75} (2012) 850;
V.~S.~Fadin,
  arXiv:1507.08756 [hep-th].
	
\bibitem{Cherednikov}	
E.~N.~Antonov, L.~N.~Lipatov, E.~A.~Kuraev and I.~O.~Cherednikov,
  Nucl.\ Phys.\ B {\bf 721} (2005) 111.
	
	
\bibitem{LightCone}	
S.~J.~Brodsky, H.~C.~Pauli and S.~S.~Pinsky,
  Phys.\ Rept.\  {\bf 301} (1998) 299;
	R.~Venugopalan,
  Lect.\ Notes Phys.\  {\bf 516} (1999) 89;
	W.~M.~Zhang,
  Chin.\ J.\ Phys.\  {\bf 32} (1994) 717.
	
	
\bibitem{Nefedov}
L.~N.~Lipatov and M.~I.~Vyazovsky,
  Nucl.\ Phys.\ B {\bf 597} (2001) 399;
  M.~Nefedov and V.~Saleev,
  ``On the one-loop calculations with Reggeized quarks,''
  Mod.\ Phys.\ Lett.\ A {\bf 32} (2017) no.40,  1750207;
  A.~V.~Karpishkov, M.~A.~Nefedov, V.~A.~Saleev and A.~V.~Shipilova,
  Phys.\ Part.\ Nucl.\  {\bf 48} (2017) no.5,  827
   [Fiz.\ Elem.\ Chast.\ Atom.\ Yadra {\bf 48} (2017) no.5 ].	
	
	
\bibitem{RegLoops}	
E.~Levin, J.~Miller and A.~Prygarin,
  Nucl.\ Phys.\ A {\bf 806} (2008) 245;
E.~Levin and A.~Prygarin,
  Eur.\ Phys.\ J.\ C {\bf 53}, 385 (2008);
 M.~A.~Braun and A.~N.~Tarasov,
  Eur.\ Phys.\ J.\ C {\bf 58} (2008) 383;	
M.~A.~Braun,
  Eur.\ Phys.\ J.\ C {\bf 63} (2009) 287;
T.~Altinoluk, A.~Kovner, M.~Lublinsky and J.~Peressutti,
  JHEP {\bf 0903} (2009) 109;		
M.~A.~Braun and A.~Tarasov,
  Eur.\ Phys.\ J.\ C {\bf 69} (2010) 75;	
S.~Bondarenko,
  Eur.\ Phys.\ J.\ C {\bf 71}, 1587 (2011); 
 R.~S.~Kolevatov, K.~G.~Boreskov and L.~V.~Bravina,
  Eur.\ Phys.\ J.\ C {\bf 71} (2011) 1757;
M.~A.~Braun and A.~N.~Tarasov,
  Phys.\ Lett.\ B {\bf 726} (2013) 300;
 T.~Altinoluk, A.~Kovner, E.~Levin and M.~Lublinsky,
  JHEP {\bf 1404} (2014) 075;
A.~E.~Bolshov, L.~V.~Bork and A.~I.~Onishchenko,
  arXiv:1802.03986 [hep-th].
	
 

\end{thebibliography}
\end{document}